\documentclass[a4paper]{jpconf}
\usepackage{graphicx}
\begin{document}
\title{Recent STAR results and prospects of $W^{-(+)}$ boson production in polarized $\vec{p}+\vec{p}$ collisions at RHIC}

\author{Joseph Seele for the STAR Collaboration}

\address{Massachusetts Institute of Technology, 77 Massachusetts Ave, Cambridge, MA 02139 USA}

\ead{seelej@mit.edu}

\begin{abstract}
In 2009, the RHIC spin physics program completed the first data taking period
of polarized $\vec{p}+\vec{p}$ collisions at $\sqrt{s} = 500$ GeV, accumulating $\sim$12 pb$^{-1}$ with $\sim$39\% beam polarization. This opens a new era in the study of the spin-flavor structure of the proton based on the production of $W^{-(+)}$ bosons. $W^{-(+)}$ bosons are produced in $\bar{u}+d (\bar{d}+u)$ collisions and can be detected through their leptonic decays, $e^-+\bar{\nu_e}(e^++\nu_e)$, where only the respective charged lepton is measured.

The discrimination of $\bar{u}+d$ and $\bar{d}+u$ quark combinations requires distinguishing the charge sign of high $p_T$ electrons and positrons, which in turn requires precise tracking information. At mid-rapidity, STAR relies on the Time Projection Chamber. At forward rapidity, new tracking capabilities will be provided by the Forward GEM Tracker, consisting of six triple-GEM detectors which are under construction. The suppression of QCD background over W boson signal events by several orders of magnitude is accomplished by using the highly segmented STAR Electromagnetic Calorimeters to impose isolation criteria suppressing jet events, and vetoing dijet events based on the measured away side energy.

The status of the STAR results on the first measurements of the cross section and single spin asymmetry for $W^{-(+)}$ boson production in polarized $\vec{p}+\vec{p}$ collisions will be presented along with a discussion of prospects involving the STAR Forward GEM Tracker.
\end{abstract}

\section{Introduction}

High-energy polarized p+p collisions at $\sqrt{s} = 500$ GeV at RHIC provide a unique way
to study the partonic spin structure of the proton. Inclusive polarized deep-inelastic scattering (DIS)
experiments have shown that only $\sim$30\% of the spin of the proton is attributable to the polarization 
of the quarks \cite{Filippone:2001ux}. However, these inclusive measurements do not discern between  the various flavours of the 
quarks and their individual contributions. Semi-inclusive DIS measurements, however,
can achieve separation of the quark and anti-quark spin contributions by flavour \cite{Airapetian:2004zf}. The extracted anti-quark polarized Parton Distribution Functions (PDFs) have sizable
uncertainties compared to the well-constrained quark and anti-quark sums \cite{deFlorian:2009vb}.

$W^{-(+)}$ bosons are produced in p+p collisions, at leading order, through $\bar{u}+d (\bar{d}+u)$ interactions at the partonic
level and can be detected through their leptonic decays. The parity-violating nature of the weak
interaction gives rise to large single beam helicity asymmetries, $A_L$, which yield a
direct and independent probe of the quark and anti-quark polarized PDFs. The single helicity asymmetry is
defined as $A_L = \left(\sigma^+-\sigma^-\right)/\left(\sigma^++\sigma^-\right)$, where $\sigma^{+(-)}$ 
refers to the cross section when the helicity of
the polarized proton beam is positive (negative). Theoretical frameworks have been developed
to describe the production of W$^\pm$ bosons and their decay leptons in polarized p+p collisions \cite{deFlorian:2010aa,Nadolsky:2003ga}.

\section{Recent Measurements}

The measurements described in these proceedings were done using the STAR detector \cite{Ackermann:2002ad} at RHIC.
The detector systems employed in this measurement are the Time Projection Chamber
(TPC), which provides tracking of charged particles in a 0.5 T solenoidal magnetic field for
pseudorapidities of $|\eta|\leq1.3$ and the Barrel and Endcap Electromagnetic Calorimeters
(BEMC,EEMC), which are lead-scintillator sampling calorimeters covering $|\eta|\leq 1$ 
and $1.09\leq\eta\leq 2$, respectively. All three detectors provide 2$\pi$ coverage in azimuthal angle, $\phi$.
The data presented in this contribution, which correspond to 13.7 $\pm$ 0.3 (stat) $\pm$ 3.1(syst) pb$^{-1}$
of sampled luminosity, were accumulated in 2009 when the first
significant dataset was collected for polarized proton collisions at a center of mass energy of $\sqrt{s} = 
500$ GeV. During this period, the beam polarizations averaged (38 $\pm$ 3)\% and (40 $\pm$ 5)\% for 
the two beams. 

In this analysis, only the $W^\pm\rightarrow e^\pm+\nu$ decay channel was considered. The electrons and positrons at mid-rapidity from these W$^\pm$ decays can be detected by a large transverse energy, $E_T$, peaked near $M_W/2$. In order to preferentially select the events containing a high energy $e^\pm$, a two-stage online BEMC trigger was required. First an event was required to satisfy a hardware
threshold corresponding to a transverse energy,  $E_T > 7.3$ GeV, in a single BEMC tower. Additionally, a software level trigger
then searched for a seed tower with $E_T > 5$ GeV and required that the maximum 2$\times$2 tower cluster
including that seed have an $E_T$ sum larger than 13 GeV. 

\begin{figure}[!t]
\begin{center}
\includegraphics[scale=1.0]{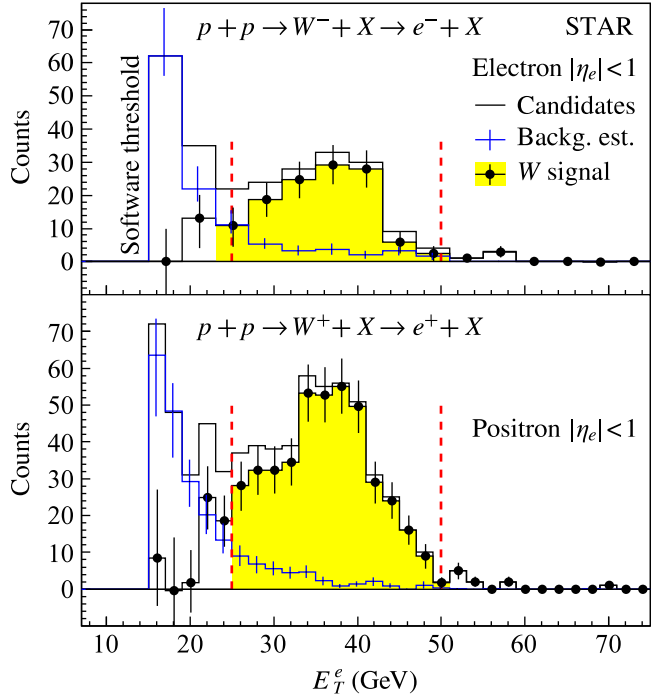}
\caption{The charge separated W$^\pm$ candidates distributions (black), the estimate of the background (blue) and the extracted signal (yellow) \cite{Aggarwal:2010vc}.}
\label{fig-jacob}
\end{center}
\end{figure}

W$^\pm$ candidate events were selected offline based on kinematical and topological differences
between leptonic $W^\pm$ decay events and QCD background (e.g. di-jets) events. $W^\pm\rightarrow e^\pm+\nu$
decay events contain
a nearly isolated $e^\pm$ and an undetected neutrino opposite in azimuth, leading to a large missing
$E_T$. The selection proceeded in three stages : $e^\pm$ candidate identification, isolation, and background
rejection. An $e^\pm$ candidate is identified as any TPC track with
$p_T > 10$ GeV/c which originated from the event vertex with $|z| < 100$ cm, where $z$ is the 
direction along the beamline from the nominal interaction point. Furthermore, it was required that this track
points to a 2$\times$2 BEMC tower cluster with an $E_T$ sum, $E^{2\times 2}_T$, greater than 15 GeV and 
whose centroid was less than 7 cm from the extrapolated track.  Next, two isolation cuts are imposed on
the candidate. First, successful candidates were required to have an excess $E_T$ in the surrounding 4$\times$4 tower cluster be less than 5\% of $E^{2\times 2}_T$. Secondly, the candidates were required to have an excess EMC tower + TPC track $E_T$ sum of less than 12\% of $E^{2\times 2}_T$ within a cone radius $R = \sqrt{(\Delta\eta)^2+(\Delta\phi)^2}= 0.7$ of the candidate. Two background rejection cuts were applied. First, the magnitude of the vector $p_T$ sum
of the $e^\pm$ candidate $p_T$ vector and the $p_T$ vectors of all the reconstructed jets with thrust axes 
outside the $R = 0.7$ cone around the candidate was required to be greater than 15 GeV/c. The jets were 
reconstructed using the standard mid-point cone algorithm used in previous STAR jet 
measurements \cite{Abelev:2006uq,Abelev:2007vt}. Second, the away-side $E_T$,  which was defined to be  the EMC + TPC $E_T$ sum over the 
full range in pseudorapidity and having $\Delta\phi > 3\pi/4$ from the $e^\pm$ candidate track, was required to be less than 30 GeV.

\begin{figure}[!t]
\hfill
\begin{minipage}[!t]{.45\textwidth}
\begin{center}  
\includegraphics[scale=0.34]{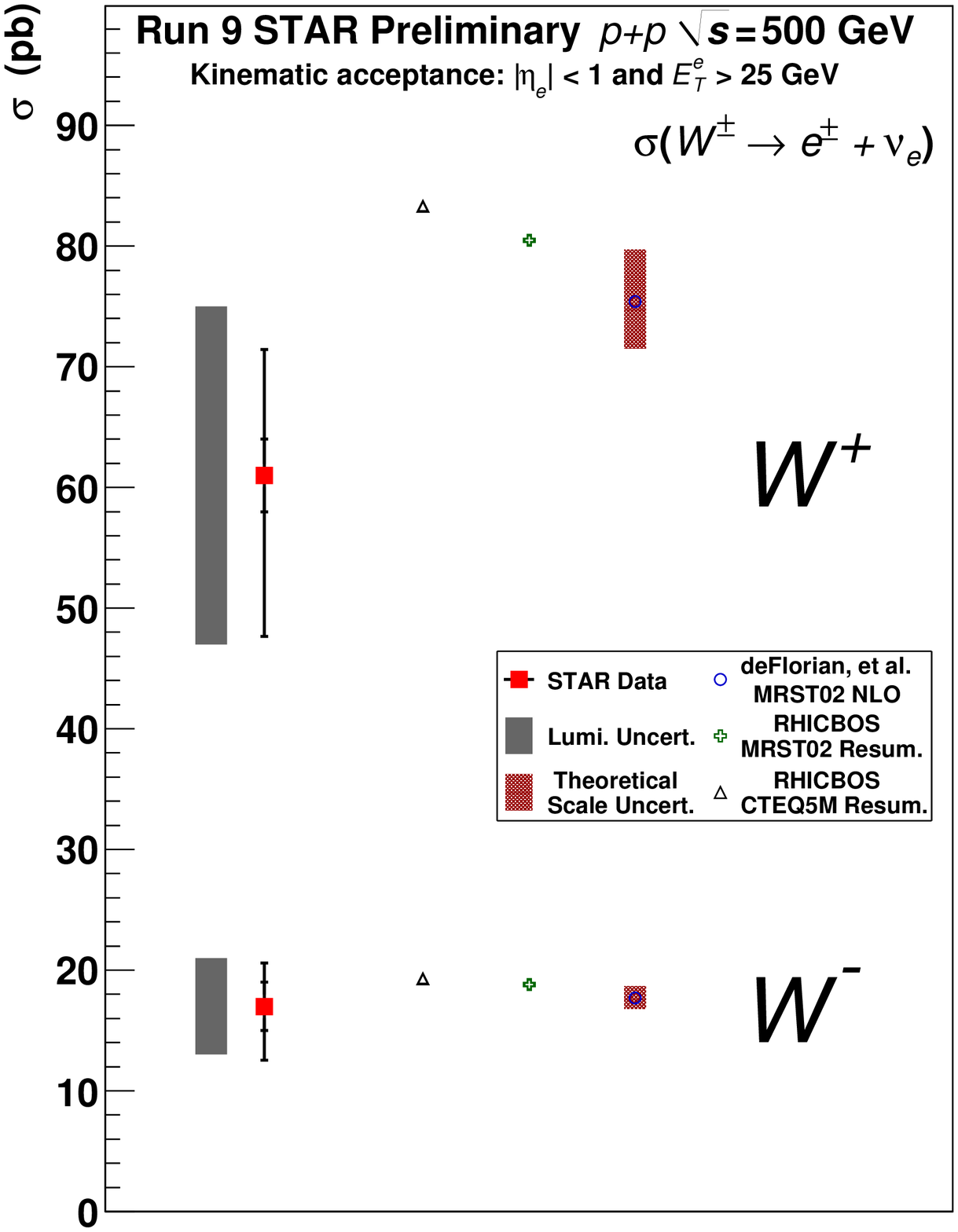}
\caption{The extracted cross sections for $W^\pm\rightarrow e^\pm+\nu$ in proton-proton collisions at $\sqrt{s}=500$ GeV. Also plotted are theoretical expectations \cite{deFlorian:2010aa,Nadolsky:2003ga} for a variety of PDFs.}
\label{fig-xsec}
\end{center}
\end{minipage}
\hfill
\begin{minipage}[!t]{.45\textwidth}
\begin{center}  
\includegraphics[scale=0.8]{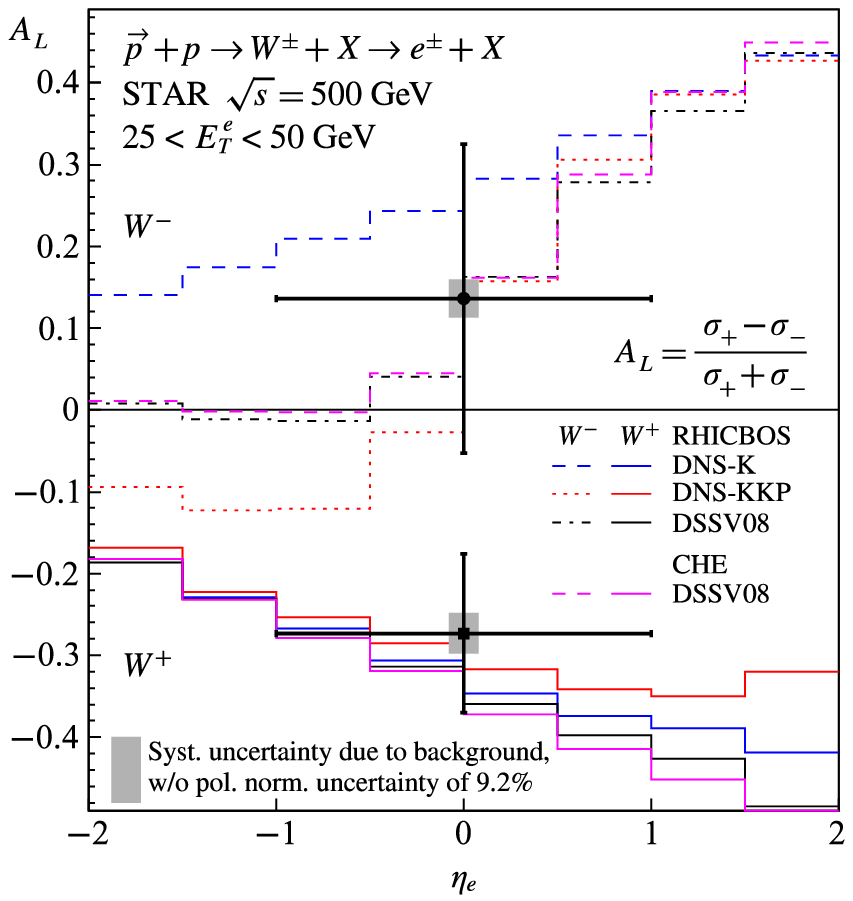}
\caption{The extracted single helicity asymmetries for $W^\pm\rightarrow e^\pm+\nu$ in polarized proton-proton collisions at $\sqrt{s}=500$ GeV  \cite{Aggarwal:2010vc}. Also plotted are theoretical expectations \cite{deFlorian:2010aa,Nadolsky:2003ga} for a variety of polarized PDFs.}
\label{fig-al}
\end{center}
\end{minipage}
\hfill
\end{figure}

After these cuts were applied, the W$^\pm$ candidates (see figure \ref{fig-jacob}) show the characteristic Jacobian 
peak at $E^{2\times 2}_T\sim M_W$/2. The amount of background remaining in the W$^\pm$ candidate sample, after 
applying the selection criteria described above, was estimated from three contributions. The first contribution was
from $W^\pm\rightarrow\tau^\pm+\nu$ decay where the $\tau^\pm$ decays semi-leptonically to an $e^\pm$ and two neutrinos. 
This background was estimated using Monte Carlo simulation. 
Another contribution estimated the impact of the missing calorimetric coverage for $-2\leq\eta\leq -1.09$. To
determine this contribution to the background, the analysis was performed a second time with
the EEMC not used as an active detector. The difference in the W$^\pm$ candidate $E^{2\times 2}_T$
distribution with and without the EEMC included in the analysis was taken to be the estimate for the missing
calorimetric coverage. The third contribution was estimated by normalizing a data-driven background
shape to the remaining W$^\pm$ candidate signal, in the $E^{2\times 2}_T$ range below 19 GeV, after the first two 
background contributions were removed. This data-driven background  shape was obtained by inverting the last 
two requirements in the W$^\pm$ candidate selection, namely by requiring that the away-side $E_T$ be greater than 30 
GeV or the magnitude of the vector $p_T$ sum be less than 15 GeV/c. The total background was then subtracted 
from the W$^\pm$ candidate spectrum to yield the W$^\pm$ signal spectrum (see figure \ref{fig-jacob}). A systematic uncertainty 
for the background estimation was determined by varying the inverted cuts used to obtain the data-driven
background shape and by varying the range where the background shape was normalized to the
remaining W$^\pm$ candidate signal.

\begin{figure}[!t]
\begin{center}
\includegraphics[scale=0.4]{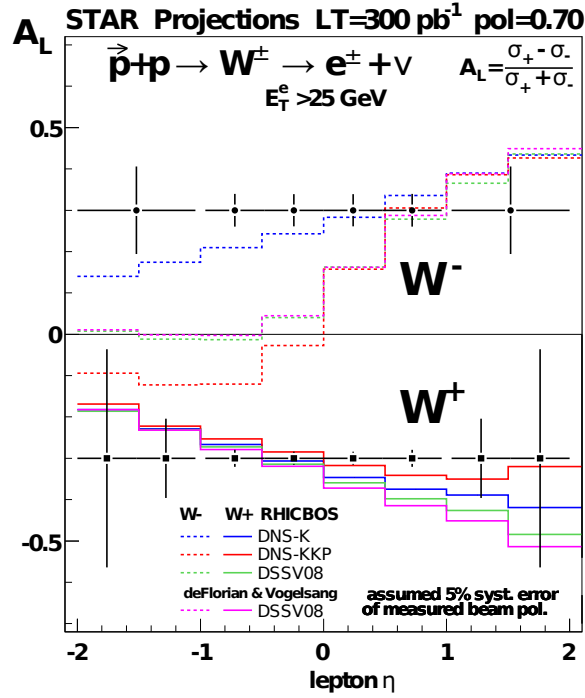}
\caption{Expected future sensitivities for the single helicity asymmetry for $W^\pm$ production in polarized proton-proton collisions at $\sqrt{s}=500$ GeV at RHIC. Also plotted are theoretical expectations \cite{deFlorian:2010aa,Nadolsky:2003ga} for a variety of polarized PDFs.}
\label{fig-al_proj}
\end{center}
\end{figure}

Preliminary results for the production cross sections and final results for the single helicity asymmetries of $W^\pm\rightarrow e^
\pm+\nu$ from events  with $|\eta|\leq 1$ and $E^{2\times 2}_T > 25$ GeV can be seen in figures 
\ref{fig-xsec} and \ref{fig-al}. The measured values are $\sigma(W^+\rightarrow e^++\nu)$ = 61 $\pm$ 3 (stat.) 
$^{+10}_{-13}$ (syst.) $\pm$14 (lumi.) pb and $\sigma(W^-\rightarrow e^-+\bar{\nu})$ = 17 $\pm$  2
(stat.) $^{+3}_{-4}$ (syst.) $\pm$ 4 (lumi.) pb. The measured asymmetries are $A^{W^+}_L = -0.33 \pm 
0.10$ (stat.) $\pm$ 0.04 (syst.) and $A^{W^-}_L = 0.18 \pm 0.19$ (stat.) $^{+0.04}_{-0.03}$ (syst.). These 
results are consistent with the theoretical calculations using the same acceptance and polarized PDFs which have been constrained by polarized DIS experiments.

\section{Future Measurements}

In the future, STAR plans to take more data at $\sqrt{s}=500$ GeV. This data will place a clean constraint on the quark and anti-quark polarized PDFs of the proton. The expected sensitivities for the single helicity asymmetry as a function of pseudorapidity can be seen in figure \ref{fig-al_proj}. The calculation in the figure assumed our demonstrated efficiency and background rejection in the mid-rapidity region while in the forward and backward rapidity regions, a worse signal to background ratio was assumed. The uncertainties are for 300 pb$^{-1}$ of sampled luminosity and 70\% average polarization which is the the goal of the longitudinal $\sqrt{s}=500$ GeV spin program at RHIC.

\section{Summary}

The STAR Collaboration has taken its first data with longitudinally polarized proton beams at $\sqrt{s}=500$ GeV in 2009. First results on the mid-rapidity cross section and single helicity asymmetry for $W^\pm$ were presented.  This opens a new era in the study of the spin-flavour structure of the proton 
based on the production of $W^{\pm}$ bosons. 

\section*{References}
\bibliography{iopart-num}

\end{document}